\documentclass[12pt,preprint]{aastex}
\begin{document}

\title{EVALUATING POTENTIAL FOR DATA ASSIMILATION IN A FLUX-TRANSPORT 
DYNAMO MODEL BY ASSESSING SENSITIVITY AND RESPONSE TO MERIDIONAL FLOW VARIATION}
\author{MAUSUMI DIKPATI}
\affil{High Altitude Observatory, National Center for Atmospheric
Research \footnote{The National Center
for Atmospheric Research is sponsored by the
National Science Foundation. },
3080 Center Green, Boulder, Colorado 80301; dikpati@ucar.edu}
\author{JEFFREY L. ANDERSON}
\affil{Data Assimilation Research Section, IMAGe, National Center 
for Atmospheric Research, 1850 Table Mesa Dr., Boulder, 
Colorado 80305; jla@ucar.edu}

\begin{abstract}

We estimate here a flux-transport dynamo model's response time to
changes in meridional flow speed. Time-variation in meridional
flow primarily determines the shape of a cycle in this class of
dynamo models. In order to simultaneously predict the shape,
amplitude and timing of a solar cycle by implementing 
an Ensemble Kalman Filter in the framework of Data Assimilation 
Research Testbed (DART), it is important to know the model's 
sensitivity to flow variation. Guided by observations we consider 
a smooth increase or decrease in meridional flow speed for a 
specified time (a few months to a few years), after which the 
flow speed comes back to the steady speed, and implement that 
time-varying meridional flow at different phases of solar cycle. 
We find that the model's response time to change in flow speed peaks  
at four to six months if the flow change lasts for one year. 
The longer the changed flow lasts, the longer the model takes to respond.
Magnetic diffusivity has no influence in model's response to flow 
variation as long as the dynamo operates in the advection-dominated 
regime. Experiments with more complex flow variations indicate
that the shape and amplitude of flow-perturbation have no influence 
in the estimate of model's response time. 

\end{abstract}


\section{Introduction}

There has been substantial interest in predicting future solar cycles for
the past forty years \citep{ohl66,ohl79}. In the current era of extensive use 
of the high atmosphere and neighboring interplanetary medium by man, such
predictions have considerable practical value. For cycles 22 and 23,
prediction methods were primarily statistical rather than dynamical. That is,
no physical laws were integrated forward in time, as is done for meteorological
and climate predictions. But for solar cycle 24 the first such cycle
prediction, which involves integrating forward in time a form of Faraday's
law of electromagnetic theory, has been made \citep{ddg06,ccj07}. 

Given the step-by-step successes of different kinematic dynamo models, 
starting from convection zone dynamos \citep{stix76}, interface dynamos 
\citep{parker93}, up to flux-transport dynamos \citep{ws91,csd95,durney95,
dc99,krs01,jetal08,gdd09} applied to the Sun, a set of kinematic flux-transport
dynamo equations was chosen to numerically integrate forward in time 
\citep{dg06}. This was analogous to what was done in the 1950s with the 
earliest weather forecast models, which were 2D latitude-longitude models. 
Kinematic flux-transport dynamo models are also 2D, but in latitude and
radius.

The kinematic dynamo equations were calibrated with solar observations,
and driven by input of observed solar magnetic data. The data input was 
continuous in time but quite simple -- a form of 'data-nudging', previously
used in early weather and climate forecast models. These calculations
simulated the relative peaks of the past cycles \citep{ddg06,ccj07}
and showed skill even when North and South hemispheres were simulated 
separately \citep{dgdg07}.

However, the predictive skill has been limited to hindcasting the 
peak-amplitude of a cycle in those calculations; for example, see
Figure 1. Here a simulated sequence of cycles derived from a tachocline 
toroidal flux integral has been superimposed on observed cycles of 
monthly smoothed
sunspot number (taken from www.sidc.be). The details within a cycle, 
such as its shape and its rise and fall patterns, have not been reproduced. 
In order to be able to predict the amplitude, timing and shape of a 
cycle simultaneously, we need to go beyond simple data-nudging 
for the entire span of integration. Updating the unknown time variations in
the dynamo ingredients in a finite interval within a cycle, say every
six to twelve months, will be required. Therefore, we need to implement 
a sophisticated data-assimilation scheme.

\begin{figure}[hbt]
\epsscale{1.0}
\plotone{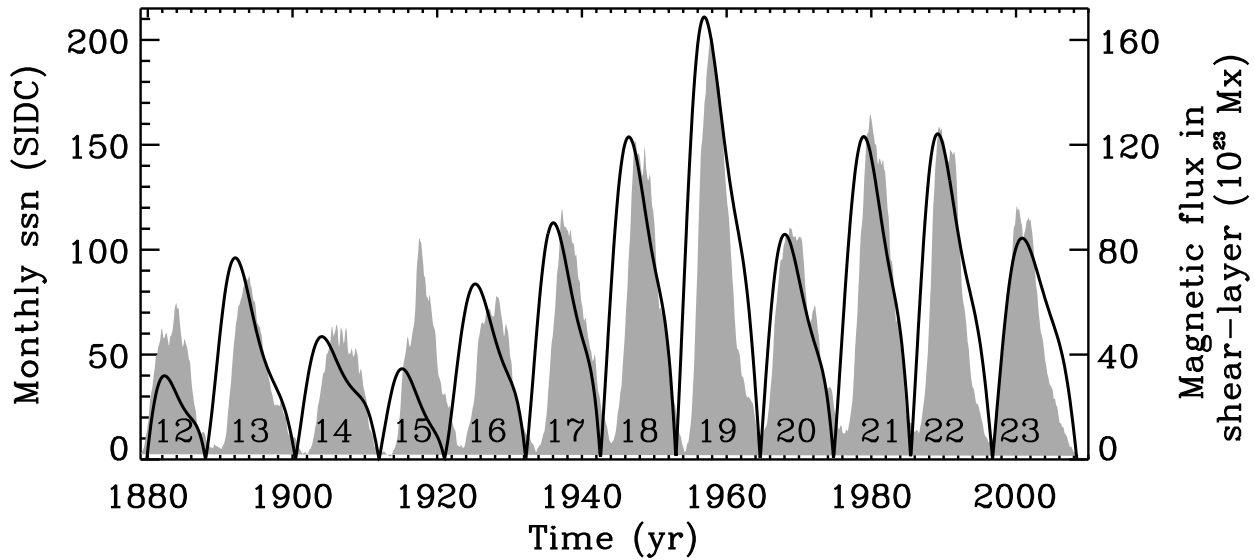}
\caption{
Gray-filled curve shows observed cycles derived from monthly smoothed
sunspot number data from the Royal Observatory of Belgium (www.sidc.be); 
superimposed on that is a simulated sequence of cycles (black curve) 
derived from total tachocline toroidal flux integral. Simple data-nudging 
for 12 successive cycles can hindcast the cycle-peaks with high skill, 
but not the details within a cycle.
}
\label{cycle-flux}
\end{figure}

Modern Earth system prediction models use sophisticated data assimilation 
methods to capture all the usable observational information about the 
system. These methods are highly developed for atmospheric and oceanic 
predictions \citep{kalnay03}, but have only recently begun to be used 
for the Sun \citep{brun07, kk09, jbt11}. With our present motivation
of simulating cycle-shape and its rise and fall patterns, a suitable
method is the Ensemble Kalman Filter (EnKF) technique in the
framework of Data Assimilation Research Testbed (DART), which has been
widely developed at the National Center for Atmospheric Research 
\citep{a01, awzh05, ac07, a09}.
Identifying the parameters which govern the spatio-temporal pattern 
of meridional flow as the so-called state vectors, we can apply an 
Ensemble Kalman Filter to these state vectors and can create an ensemble 
of time-varying magnetic fields by advancing our dynamo model. A built-in 
Monte Carlo step within DART selects the simulation closest to 
observation after each specified time advancement of the model within 
a solar cycle. If we advance the model sequentially over an entire solar 
cycle and determine the rise and fall patterns of that cycle that match 
best with observations, then we can construct the spatio-temporal pattern 
of meridional flow for an entire cycle. Similarly, if we know the
spatio-temporal variations of meridional flow in a cycle, we can
predict the shape, rise and fall patterns of that cycle.

\clearpage
\begin{figure}[hbt]
\epsscale{1.0}
\plotone{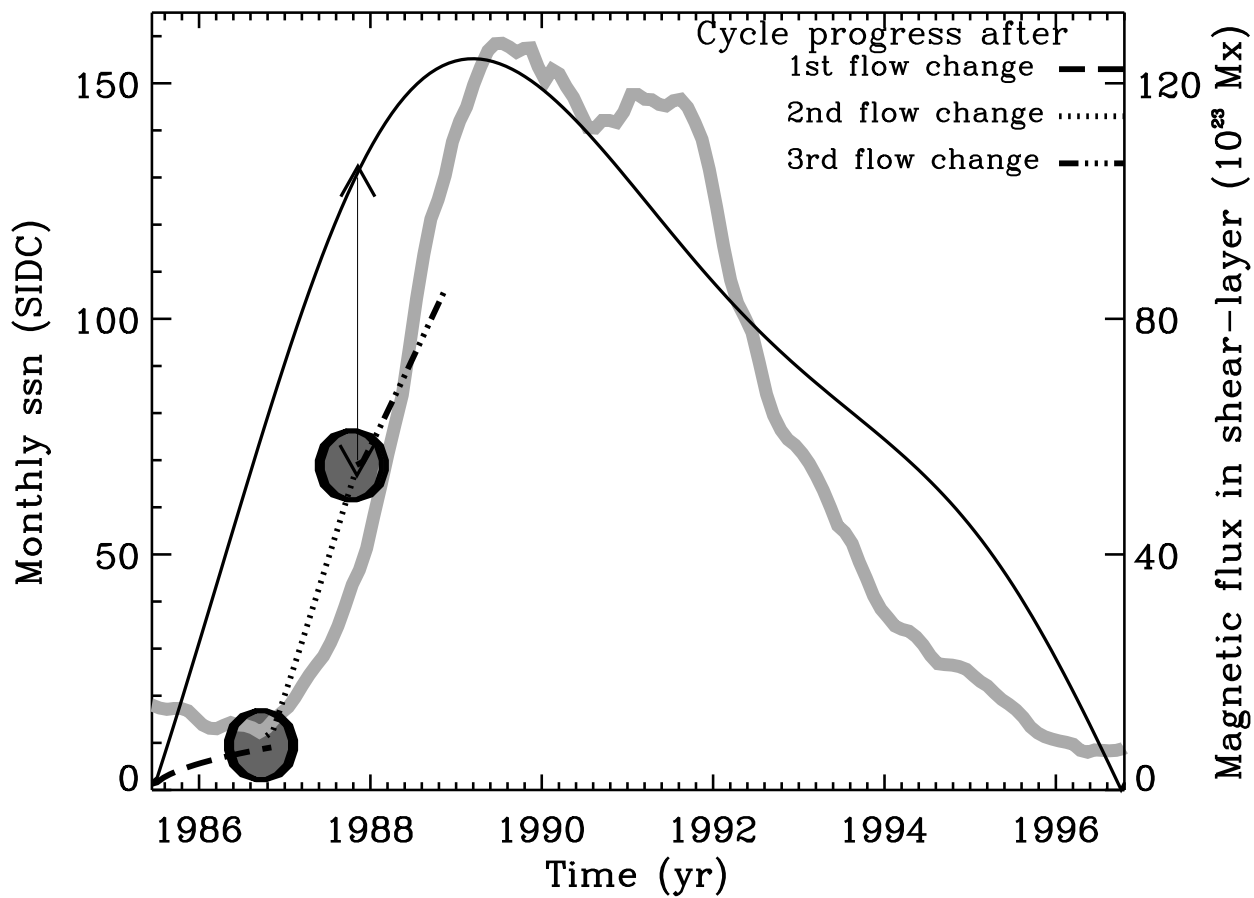}
\caption{
Thick gray curve represents an observed cycle based on monthly 
smoothed sunspot number; superimposed on it is a corresponding simulated
cycle measured by its tachocline flux integral. Successive dashed, dotted and 
dash-triple-dotted curves describe a sequential data-assimilation method
with three successive flow variations during the forward time-advancement
of the dynamo model. Meridional flow-speed is adjusted in the model 
after each specified time-interval during a simulation-run. Combining
the curves -- dashed, dotted and dash-triple-dotted curves and those 
beyond -- one member of the Ensemble can be constructed.
}
\label{schem-diagram}
\end{figure} 
\clearpage

Figure 2 describes how the cycle shape can change when the meridional
flow speed changes after a specified time. From Figure 1, the observed 
cycle 22, derived from monthly smoothed sunspot number data,
has been plotted in Figure 2 as a thick gray curve, and the simulated 
cycle (black curve) has been superimposed on it. As an example, Figure 2 
clearly reveals that the simulated cycle's rise and fall patterns do 
not match with that of the observed one, because the simulation was 
performed with a steady meridional flow (see \citet{dg06} for details). 
Spatio-temporal variations in meridional flow were not known before the
1980's. Very recently the surface flow-patterns as functions of latitude 
and time are being detected \citep{ulrich10} (see also 
\citet{grbd11}). So, running the dynamo simulation, including updating 
the meridional flow speed after a specified time, and checking how the 
model-output compares with the observation, we can construct members of 
the Ensemble for a desired solar cycle (e.g. cycle 22, the example in 
Figure 2) and also the spatio-temporal variations of the meridional 
flow for that cycle.  

From our prior knowledge of the properties of flux-transport dynamos,
we can make certain guesses about how the simulated cycle's phase will 
progress as the meridional flow varies. For example, near the beginning
of the cycle (see Figure 2), we see that the simulated cycle-phase
has progressed much faster than the observed one. Among various
possibilities, a decrease in the flow-speed would make the cycle-phase
progress more slowly, as shown by the dashed line. After a specified time,
by comparing model-output with observations, the flow-speed can be updated
again, with the aim of securing the closest possible match of the model
cycle phase with the observed phase.   

In this regard, it is first necessary to know which ingredients in the
model determine the shape of a cycle, how those ingredients vary with time
and how the model responds to their time-variation. Previous flux-transport
dynamo studies \citep{dc99} indicate that the meridional circulation is
the key ingredient that determines the timing and shape of a cycle.
We know from observations that the meridional circulations vary 
substantially with time for both the Earth and the Sun. 
Perturbations in meridional flow are likely to be particularly 
important, since from considerations of mass conservation these 
perturbations are likely to be felt rather quickly throughout the 
model domain.

However, to make effective use of all the observational data available
to predict future behavior of the Sun or Earth system, it is important
to determine the 'response time' of these systems to perturbations
of various types and durations occurring at various places within the
system. This has been studied extensively for ocean circulation and climate
systems \citep{neelin11, aa08, s04, srs06, sm99, hch80, bh89},
but not yet for the solar case. There does not appear to be a single
'response time', but rather a continuous range of times, beginning soon
after the time the meridional circulation starts changing, continuing
through a time of 'peak' response, followed thereafter by a time
of declining further response. The time of peak response itself
varies considerably, depending on the duration and location of the
perturbation introduced, as well as the range of timescales inherent in
the physics of the model. The model may respond quite differently to a
perturbation of long duration compared to one of very short duration (for a
discussion of this point in the case of climate systems, see \citet{neelin11}
section 6.8).

Comparison of the solar cycle 
prediction problem to that for the prediction of the Earth's climate system 
is particularly instructive. It is well known \citep{neelin11} that Earth's 
climate and its variations in time are determined by interactions among the 
various major components of the Earth 'system' (atmosphere, oceans, land 
surface, polar ice caps) as well as by external solar forcing. These 
interactions and forcings occur on a wide range of timescales, from days to 
centuries and millennia \citep{jbl00,dmc07,wdc00,ffsw06,jil11,cm05}. This is 
analogous to, but even more complex than, the Sun's 'climate' system, or the 
solar dynamo, in which timescales of days (for emergence of new magnetic flux 
in active regions) to years (for variations in the Sun's 'conveyor belt' or 
meridional circulation) to a decade or two (for transport of magnetic flux to 
the bottom of the convection zone) and beyond (for the envelope of the solar 
cycle and the occurrence of 'Maunder minima") are prominent.

In the Earth system, with the exception of the surface mixing layer, 
the timescales in the ocean are much longer than those in the atmosphere. 
They range from months to years for the layer above the thermocline, 
to decades to millennia for the deeper ocean. It follows that to 
predict changes in climate on timescales longer than a few weeks, the
dynamics and thermodynamics of the ocean must be included in the models 
that are used \citep{neelin11,czd86}. Analogously, to simulate solar cycle 
properties requires models that capture the MHD of the whole solar convection 
zone, for which the timescales are much longer than for emergence 
of new magnetic flux at the photosphere. Mean field flux-transport
dynamo models are among the simplest that do this.

In both the Earth and Sun meridional circulation plays a critical 
role in determining behavior of the respective systems on longer 
timescales \citep{w02,getal05,dc99,ddg06,dg06}. The closely related 
'signal storage' capacity or memory in both systems is particularly important. 
In the ocean the memory of past temperature anomalies at the ocean-atmosphere 
interface is \citep{czd86} retained in the deeper ocean, brought there, and 
later back to the surface of the ocean, by meridional flow. Similarly, the 
memory of past photospheric magnetic flux patterns is retained deep in the 
convection zone, brought there by the predominantly inward meridional 
flow at high latitudes.

This memory provides the basis for prediction of changes in the climate 
of the Earth and Sun on timescales of years to a decade or two. For the 
Earth, these predictions are focused on El Nino and La Nina events, 
as well as associated extratropical changes \citep{czd86,metal09}. For the 
Sun, the focus is on predicting how certain global properties of solar cycles, 
such as peak amplitude, duration  and shape differ from one cycle to the next.
This paper begins the process of developing a more sophisticated data assimilation
scheme by estimating a flux-transport dynamo model's response time to variations 
in one of its crucial ingredients, the meridional flow. The results will be
important for achieving skill in predicting details within a cycle, the 
cycle-shape, its rise and fall patterns.

\section{Steps Towards Building Sequential Data Assimilation 
Procedure}

In order to implement the EnKF sequential data assimilation scheme
for the purpose of successfully predicting cycle-shape, we need essentially
two steps: (i) develop the implementation procedure of the scheme 
in the case of a flux-transport dynamo model, (ii) determine precisely 
the response of that dynamo model to changes in ingredients that govern
the cycle-shape. In this particular class of dynamo models,
the spatio-temporal variations in meridional flow are the most important 
ingredients for creating the variation in the progress of a cycle's phase
\citep{dgdu10}, and hence the shape of that cycle. Therefore we wish to
build a data assimilation scheme for our dynamo model that makes use of
available data on variations in meridional flow on the Sun. In order to do an 
ensemble of simulations that optimizes use of the information contained
in the meridional flow data, we must first determine the 
'response time' of the dynamo model to changes in meridional flow. This will 
give us essential guidance about, for example, what time interval over which 
to integrate the dynamo equations before adjusting the flow. Each segment
of the model-run using sequential data assimilation should be long enough to
allow the physical system to feel the change, but not so long as to degrade the
time resolution of the simulated amplitude compared to the observed one.
We explain how we do the response time calculation in section 2.3. The results
of these calculations are shown in section 3.

\subsection{Dynamo equations}

The dynamo equations we use here are the standard ones for flux-transport 
dynamos \citep{dc99}, given by 

$${\partial B \over \partial t} = -{1 \over r} \left[{\partial \over
\partial r}(r {u_r} {B}) + {\partial \over \partial
\theta}({u_{\theta}}
{B}) \right]+ r\sin\theta ({{\bf B}_p} .\nabla)\Omega \hspace{6cm}
$$
$$ \hspace{6.0cm}
-{\bf{\hat e}}_{\phi}\thinspace . \thinspace \left[
\nabla\eta\times\nabla\times {B} {\bf{\hat
e}}_{\phi}\right]+\eta({\nabla}^2
-{1 \over r^2 \sin^2 \theta}){B},\eqno(1)$$
$${\partial A \over \partial t} = -{1 \over r\sin\theta}({\bf u} .\nabla) 
(r\sin\theta A) + \eta\left({\nabla}^2 - {1 \over r^2 \sin^2 \theta}\right) A
+{S(r, \theta) B \over 1+{(B/B_0)}^2}+ {\alpha B \over 1+{(B/B_0)}^2}. \eqno(2)$$
 
\noindent
in which, the notations have usual meaning: $A$ and $B$ are
respectively the poloidal filed vector potential and toroidal fields, $u_r$
and $u_{\theta}$ are $r$ and $\theta$ components of meridional flow, $\Omega$
the differential rotation, $\eta$ the diffusivity, $S$ the surface poloidal
source (works as a nonlocal $\alpha$-effect on toroidal fluxtubes risen
to the surface from the tachocline) and $\alpha$ the tachocline 
$\alpha$-effect. $B_0$ is the quenching field strength, which may or
may not be the same for the nonlocal and local poloidal field sources; 
however, in this calculation the value of $B_0$ is the same for both
$\alpha$-effects.

\subsection{Estimating Model's Response Time to Flow-Change} 

Flux-transport dynamo simulations with steady flow revealed that the
dynamo cycle period is inversely proportional to the flow-speed \citep{ws91,
dc99}. This result was obtained by changing the flow speed from a previously 
set speed and, assuming that it will remain steady for several cycles, the 
dynamo was relaxed for about 4 or 5 cycles to obtain a new, saturated 
solution. However, observations \citep{ulrich10,gbs10} indicate 
that the flow speed varies within a cycle in a time shorter than 11 years. 
In that situation, the phase of the cycle should change accordingly, 
and in turn, determine the shape of that cycle. 

In order to estimate the response time of the model to a variation 
in flow, we will be using a calculation of lag-correlation between
flow change and cycle's phase-change. Lag-correlation has been
used in the calculation of response time of oceanic and atmospheric 
models to the variation of their ingredients \citep{yasunari90,bb98,
snm05,wkp06}. Correlation coefficients in general give a measure of 
linear association between two variables. Lagged 
correlations are obtained by correlating a lagged dataset (cycle's
phase change in our case) with another unlagged dataset (flow change) 
using the Pearson method. Lagged data are computed by shifting data 
by a certain unit of time, either forward or backward. In our case, we
will choose the unit of time to be 15 days. This means we will 
shift forward the lagged data, the cycle's phase-change, by multiples 
of 15 days with respect to flow-change data. We identify the forward 
lagged-time that exhibits the highest, positive correlation 
coefficient between the flow-change and cycle's phase change with the 
time of peak response of our dynamo model to flow-change.

Although the lag-correlation methods have been widely used in 
the context of atmospheric and oceanic models' response time 
calculations, it has not been used so widely in the solar models,
and, to the best of our knowledge, never in the solar dynamo models.
Hence a schematic diagram is presented in Figure 3 to describe
the lag-correlation we will use in this paper. In all frames of
Figure 3, $v_0$ denotes the amplitude of the steady flow, $v_1$,
$v_2 ... v_{10}$,  the amplitudes of the time-varying flow at 
different times $t_1$, $t_2 .... t_{10}$. In each frame of Figure 3,
the first column denotes the times at which the flow has been changed, 
the second column the change in flow-speed with respect to the steady 
flow-speed, and the third column the phase difference between the two 
simulated cycles (the cycle with time-varying flow and that with steady 
flow). In order to obtain a lagged correlation with zero, $t_1$, 
$t_2, ....$ time-lag, the second column quantities are correlated 
with the third column quantities staggering in the way indicated
by the arrows. 

In order to carry out the lag correlation analysis in the present 
context, we must define the phase change of a cycle.
It is known from previous 
studies that the increase (decrease) in meridional flow speed makes 
the cycle progress faster (slower). The change in cycle-phase caused 
by flow-variation should be reflected in a simulated cycle as, 
(i) an amplitude change at a specified time compared to the 
cycle-amplitude for a steady flow case at that time or, (ii) a 
change in cycle's progress-time due to flow-variation to achieve a 
specified cycle-amplitude. Here we will use the first definition.

\clearpage
\begin{figure}[hbt]
\epsscale{0.5}
\plotone{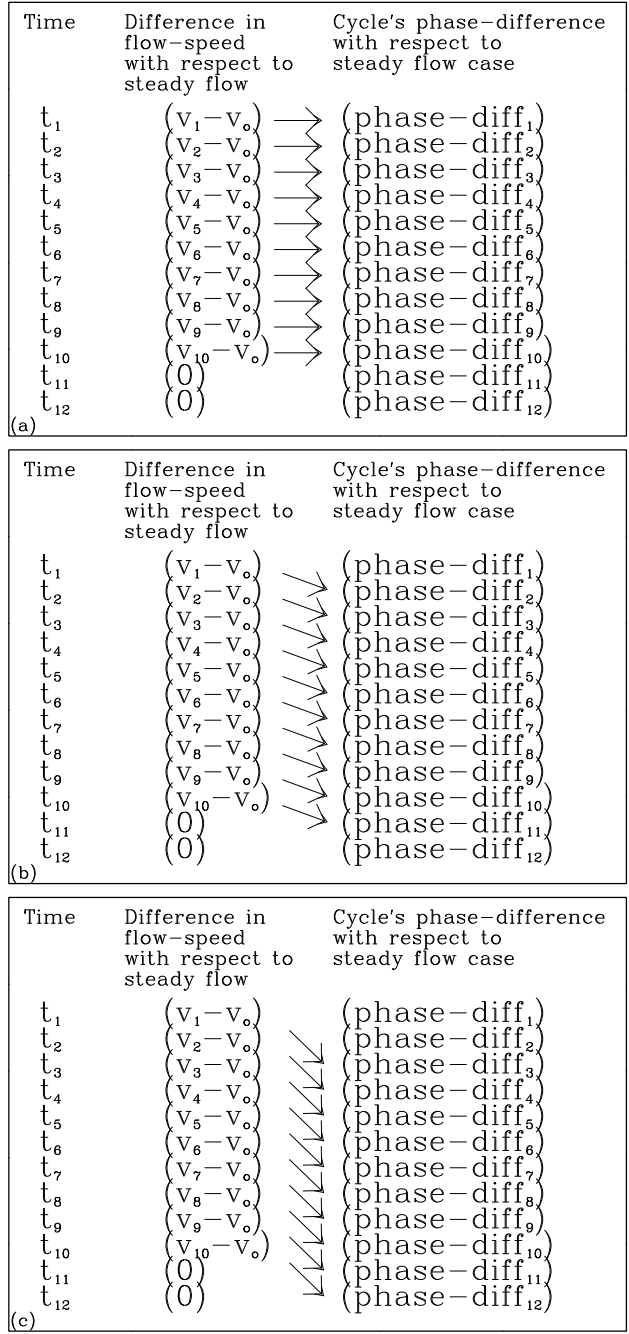}
\caption{Frames a, b and c describe how lag correlations between
flow-speed-changes (with respect to steady speed) and cycle-phase-changes 
(with respect to that with steady-flow case) are calculated. Frame (a)
has no lag between flow-speed-changes and cycle-phase-changes, frame (b)
has a $t_1$ lag (actually lead in our case here) and frame (c) has
a $t_2$ time-lag (lead here) with respect to the initial time ($t_0=0$)
when the flow-change starts.
}
\label{lag-correlation-schematic}
\end{figure}
\clearpage

Returning to Figure 2, we illustrate graphically what the
phase change is according to definition (i). The solid, black curve 
in Figure 2 represents a simulated cycle with a steady meridional
flow. Implementation of flow-change for a specified time would lead
to change in the progress of the cycle, as shown schematically
in dashed, dotted and dash-triple-dotted curves. At a specified
time, the difference in cycle-amplitudes for steady and time-varying
flow (respectively the solid, black curve and dotted curve) is shown 
by a two-sided arrow, which essentially represents the cycle 
phase-change as defined in (i). Note that the phase change is 
negative in this case.

Since we do not have observational guidance about variations in
radial of meridional circulation with time,
we will consider a fixed, single-cell meridional flow profile as 
used by \citet{dikpati11}, and vary the poleward surface flow-speed 
only. From mass conservation, this will cause instantaneously
a proportional speed change everywhere, namely for the flow sinking near
the pole, returning equatorward at the bottom of the convection
zone and upwelling near the equator. Furthermore, we will use a 
sinusoidal variation in speed with time ($\sin(\pi\,t/2\,\Delta\tau)$)
during a time-span ($\Delta\tau$) of 3 - 24 months within a cycle's
rising, peak or declining phase. This means the speed increases 
during the first half of the time-span and decreases during the 
rest, or vice versa. We will also perform a few experiments
with more complex flow profiles. Moreover, due to lack of observational
information about flow below the surface, we will consider meridional
flow profile constrained by mass-conservation in the entire dynamo
domain. Thus the flow will be a streamlined flow behaving in a self-similar
fashion, namely the percentage change in the amplitude of the flow
at a certain point at the surface will reflect the same percentage
change at all points of the fluid.

Different forms of speed variation with time can be considered,
but a step-function is not a good choice for studying response
time of a model to any ingredient variation. For some very idealized 
physical systems, a step-function is sometimes used to represent the 
change in the physical parameters for which a system response is 
calculated. In all the work on response times in geophysical fluid 
systems cited above, only one, \citep{hch80} tried a step-function 
for the system to respond to (in their case involving
deep ocean heat storage and its response to climatic forcing).
They concluded that while some properties of response can be
studied using a step-function, it is physically quite unrealistic,
so they did not base their primary results on that form of change.
Instead, as stated in \citet{hch80}, they used smooth, quasi periodic
variations, as many other investigations have done, because
they are much closer to the underlying physics and what is observed. 
The same is true for the solar meridional circulation. We therefore 
take smoothly varying changes in meridional circulation, of durations 
suggested by reference to the observational analyses of \citet{ulrich10}, 
rather than sudden step-function like changes.

Simulated solar cycles can be constructed in many ways. The cyclic 
variation of tachocline toroidal magnetic field at a selected latitude 
can be extracted (see \citet{cd00}), because that is the spot-producing
field. \citet{dg06} used a tachocline toroidal flux integral within
sunspot latitudes (equator to $45^{\circ}$) to construct theoretical
solar cycle for comparing with observed spot-area cycle. Here we will 
use three different measures for the simulated cycles: (i) tachocline 
toroidal field ($B_{\phi}$) at $15^{\circ}$ latitude, (ii) $B_{\phi}$
at $60^{\circ}$, (iii) total tachocline toroidal flux integrated from
the equator to pole. 

\section{Results}

We calculate the time of peak response of a flux-transport dynamo model for
which all the settings of the dynamo-ingredients, the differential
rotation, meridional circulation, surface and bottom $\alpha$-effects,
magnetic diffusivity and $\alpha$-quenching, are as used in \citet{dgdu10}
(see also \citet{dikpati11} for dimensional and non-dimensional parameter
values for meridional circulation). We solve the dynamo equations in the 
northern hemisphere using a single-cell meridional flow profile.    

\subsection{Model's response to flow variation with 9 months' duration}

As mentioned in \S2.3, we consider a sinusoidal variation in meridional
flow speed with a maximum amplitude of 50\% of the steady flow. We first
present the results in Figures 4-7, for the flow variation that 
lasts for 9 months ($\Delta\tau=9$ months). These figures respectively
show the results for three distinct cycle-proxies for simulations of
the same sequence of cycles: tachocline toroidal field $B_{\phi}$ at 
$15^{\circ}$ (Figures 4 and 5), $B_{\phi}$ at $60^{\circ}$ (Figure 6) and the 
total tachocline toroidal flux integral (Figure 7).

\clearpage
\begin{figure}[hbt]
\epsscale{0.85}
\plotone{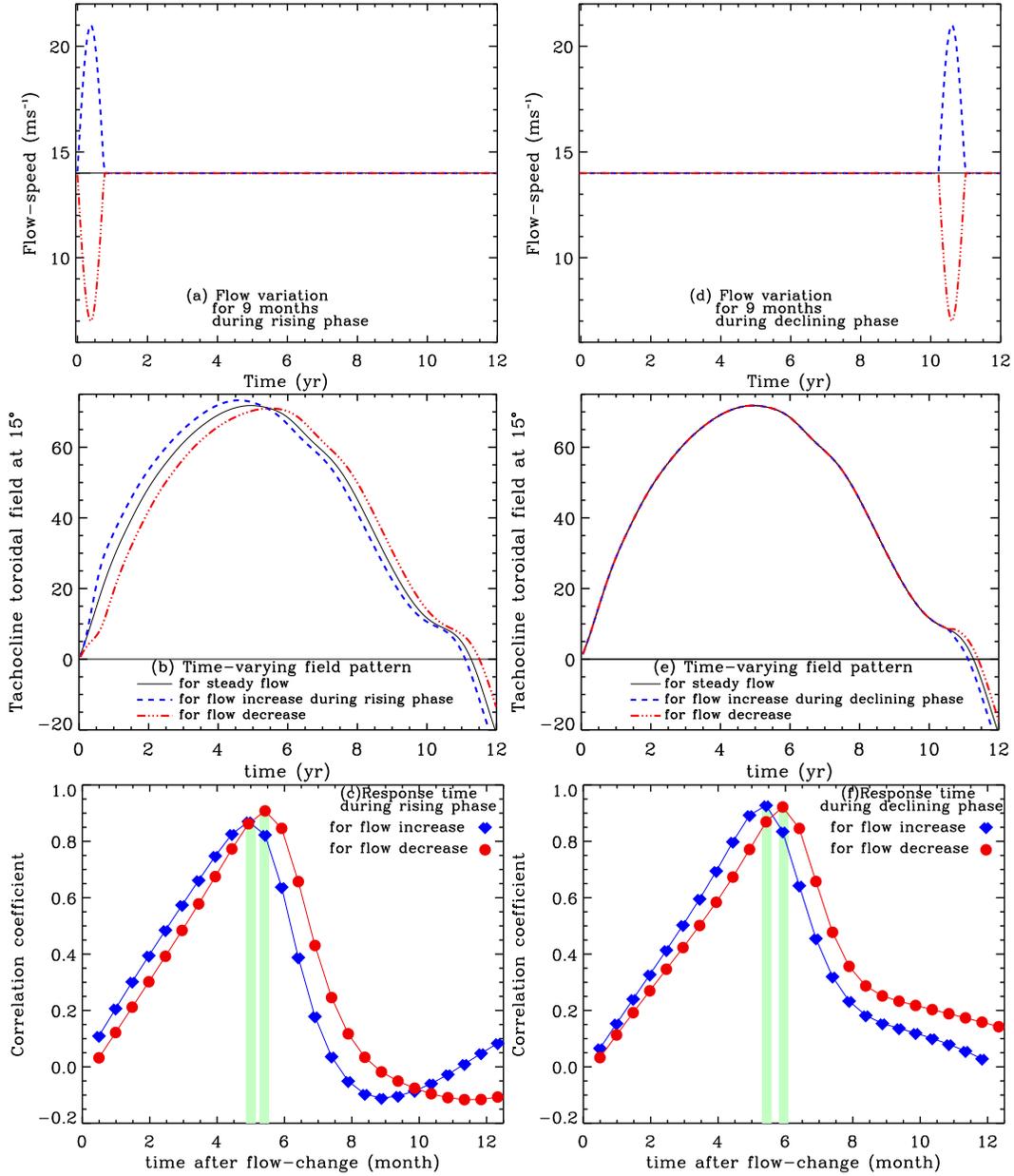}
\caption{
Top and middle frames in left column show respectively variation in meridional
flow-speed for 9 months during the rising phase of a cycle (frame a) and
cyclic toroidal magnetic field at $15^{\circ}$ latitude for steady (black)
and time-varying flows (blue and red) (frame b). Bottom frame (c) shows how
change in cycle-phase correlates with change in flow-speed as function of
time (blue-diamonds are for flow increase with respect to steady flow and 
red-filled circles are for flow decrease). The time at which cycle-phase-change 
has the highest correlation with flow-speed-change is a measure of the model's 
time of peak response to flow variation. Frames d, e, f show the same 
information as in frames a,b,c respectively, for variation in flow speed during
the declining phase of the cycle.
}
\label{field15_response}
\end{figure} 
\clearpage

\begin{figure}[hbt]
\epsscale{0.7}
\plotone{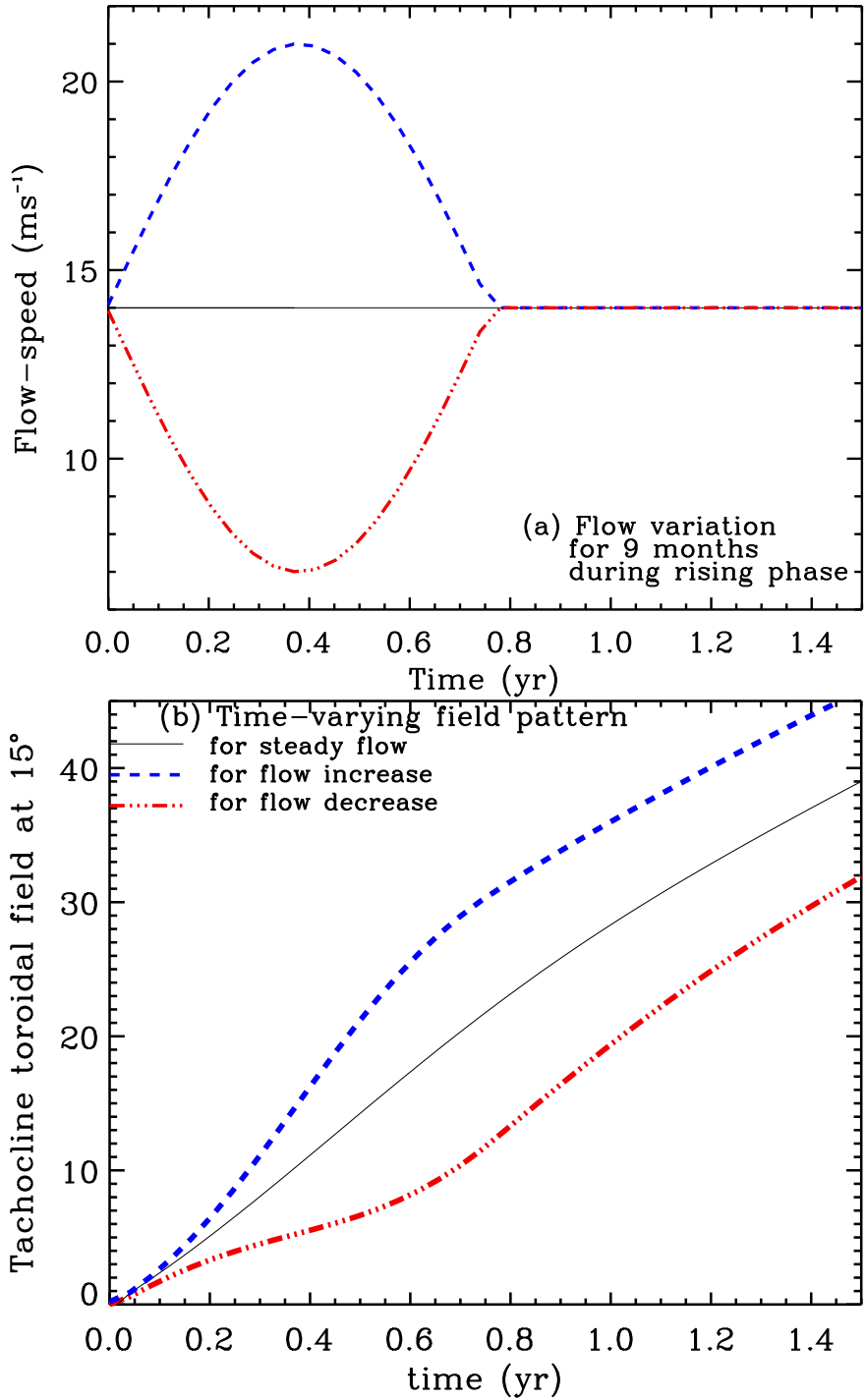}
\caption{Enlargement of first 1.5 years of response of tachocline toroidal
field at $15^{\circ}$, for the case shown in Figure 4.
}
\label{enlargement}
\end{figure}
\clearpage

$B_{\phi}|_{{\rm at} 15^{\circ}}$ should be a good proxy of the sunspot 
cycle because $15^{\circ}$ is approximately the peak sunspot latitude. 
This proxy has been used in the  past in flux-transport dynamo simulations 
(see, for example, \citet{cd00}). The three frames in the left column 
of Figure 4 show the result for flow variation that occurs during the 
rising phase of the cycle (Figure 4a), in the form of an increase 
(blue curve) or decrease (red curve) with respect to the steady flow 
(black curve). In Figure 4b we display the profiles of toroidal field 
$B_{\phi}$ as functions of time; black curve denotes the cycle with 
steady flow and blue and red curves represent that with time-varying 
flow as shown in Figure 4a. The bottom frame (Figure 4c) shows the 
lag-correlation coefficients, plotted in blue diamonds for flow 
increase and in red circles for flow decrease. The right column 
(Figures 4d-f) shows the analogous plot when the flow change occurs 
during the declining phase of the cycle.  

The results seen in Figures 4 and 6 are in many ways quite similar, implying 
that they do not depend significantly on the choice of latitude for the proxy 
toroidal field. In both cases, for meridional flow perturbations in both 
ascending and descending phases, the cycle advances in phase faster when
the meridional flow is increased temporarily, and slower when it is decreased
(Figures 4b,e; 6b,e). This effect is to be expected, given that in 
flux-transport dynamos, the cycle period is largely determined by the 
meridional flow speed \citep{dc99}.

The lag-correlation coefficients in Figures 4c, 4f, 6c, 6f all start 
with a low value at the beginning of flow-variation, slowly increase with 
the increase in time-lag between the cycle-phase and flow-change, reach
a peak and then decline rapidly. From the lag time of the occurrence of 
the peak, we estimate the time of peak response of the model to the flow change 
to be 5 to 6 months for both proxies. But from Figure 5 we can see that the
model slowly starts responding to the change in meridional flow within 
one month after the meridional circulation starts to change. So we are 
measuring the time of peak response; the system starts to respond almost
immediately.

Figures 4 and 6 reveal that the peak in lag-correlation coefficient for the 
case with flow-increase (blue diamonds) always occurs a few months before that 
with flow-decrease (red circles). This means that the model responds a little 
faster to an increase than a decrease in flow speed. We speculate that this 
difference occurs because when the speed is increased at every point in the 
domain, the 'signal' of change is transmitted faster to neighboring points, 
making it possible for the whole system to adjust faster to the change. The
opposite occurs when the flow is decreased everywhere.

\clearpage
\begin{figure}[hbt]
\epsscale{0.95}
\plotone{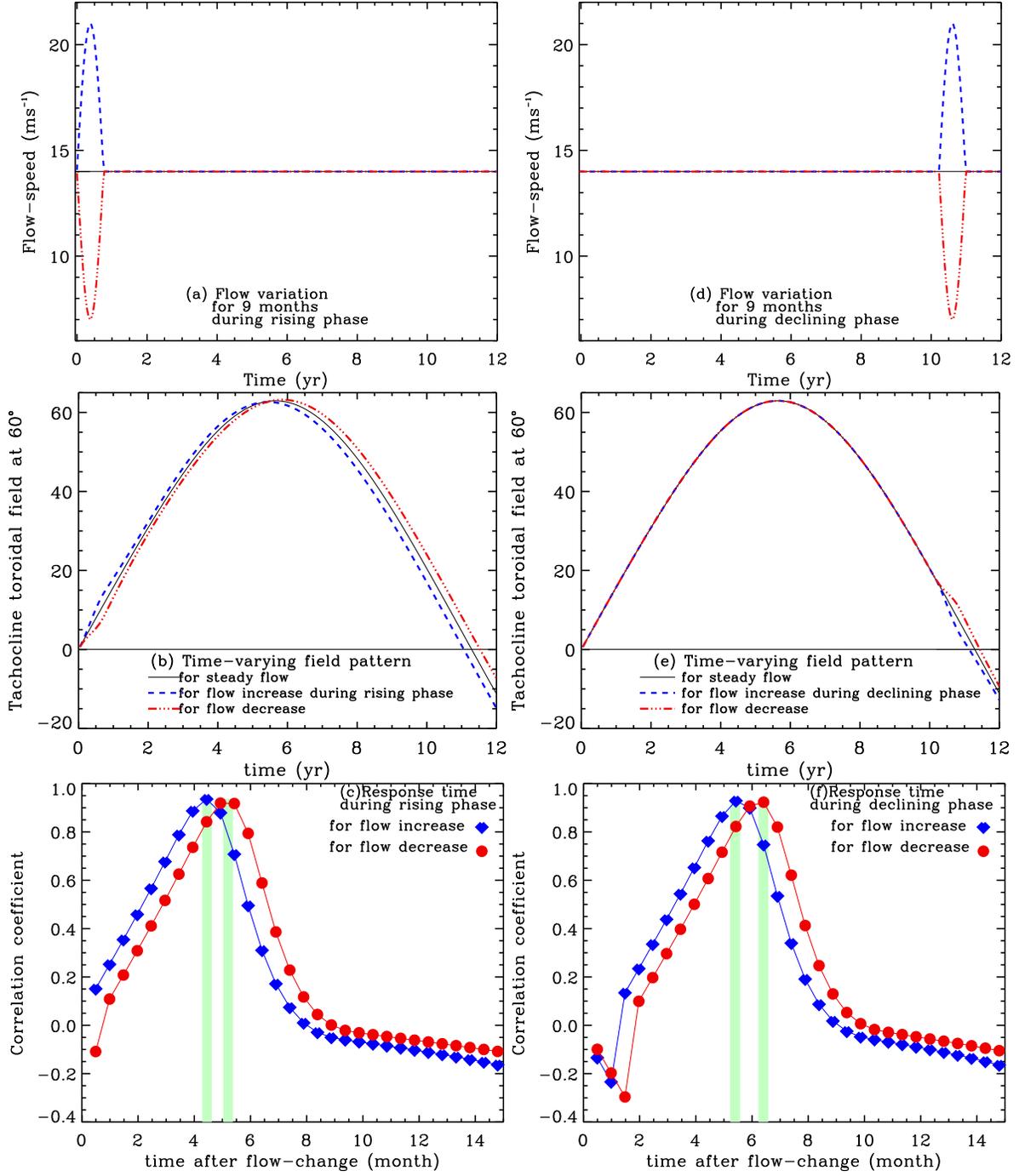}
\caption{
Same as in Figure 4, but for the tachocline toroidal field at $60^{\circ}$.
}
\label{field60_response}
\end{figure}
\clearpage

\begin{figure}[hbt]
\epsscale{0.95}
\plotone{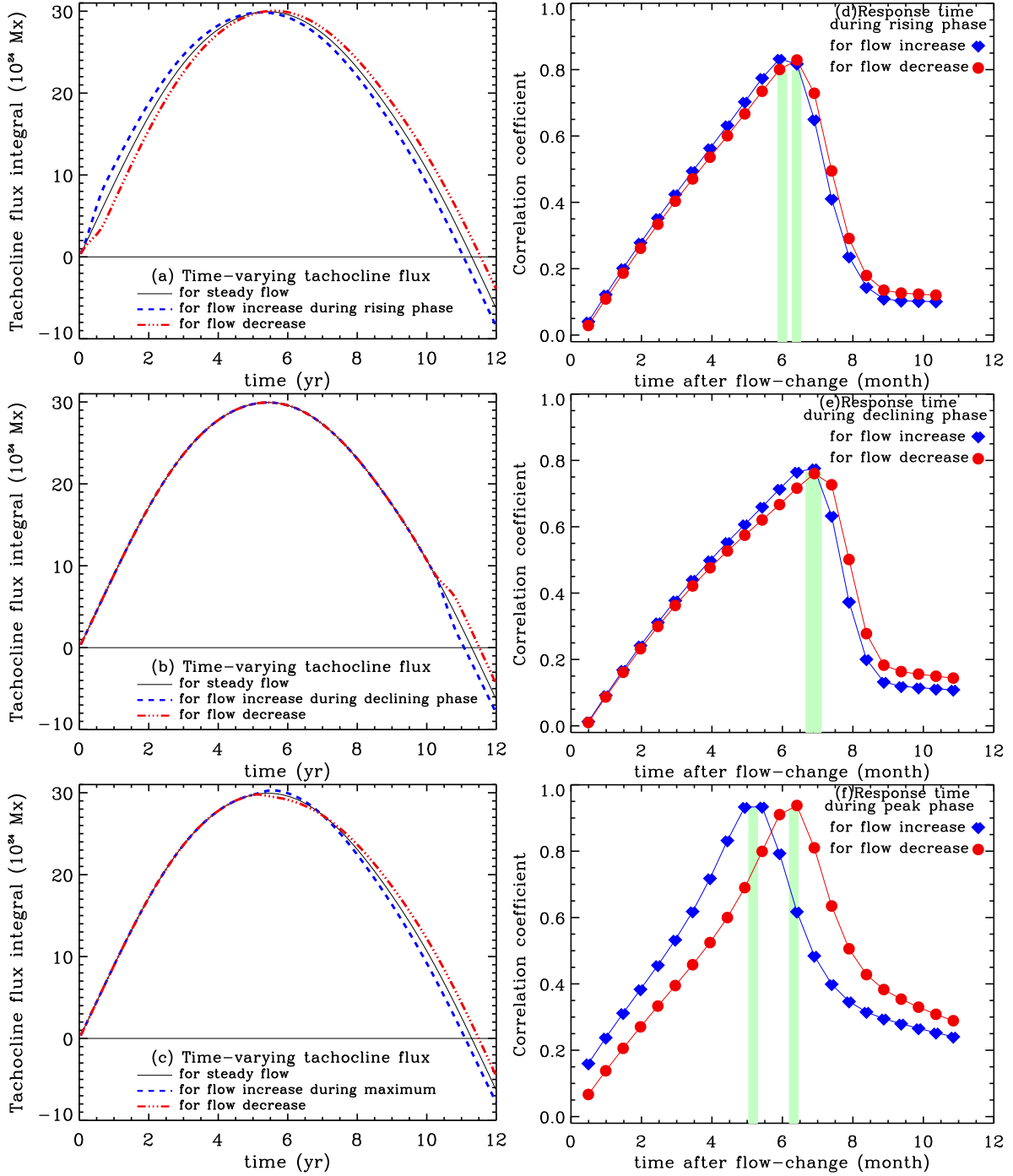}
\caption{
Three frames in left column present changes in cycle pattern
for total tachocline toroidal flux, due to changes in flow during
rising phase (frame a), declining phase (frame b) and cycle-maximum
phase (frame c). Frames (d-f) in right column present the
corresponding lag-correlation coefficients. 
}
\label{totalflux_response}
\end{figure}
\clearpage

Comparing Figures 4c with 4f and 6c with 6f, we find that the model's 
time of peak response to the flow-change is slightly longer (by about a month) 
in the declining phase than in the rising phase. It is less clear what 
causes this difference, but it may have to do with the fact that during 
the rising phase, toroidal fields are being amplified without change 
of sign, while in the late declining phase the toroidal fields are 
declining and going through a sign change.

The correlation also has a substantial time duration (a few months),
rather than a spike, indicating that the model responds over a range
of time-scales, whose peak is about 5 months in this case. This 
is what we should expect in a system in which advection by meridional
circulation and diffusion are competing with each other; in other
words the magnetic fields are only partially frozen in the plasma,
unlike an ideal MHD case where we may find a delta function type
response of the model.

Figure 7 shows the profiles of lag-correlations for the total tachocline
flux integral for meridional flow perturbations of nine month's duration for
three cycle phases: ascending and declining phases as seen in Figures 4 and 6, 
and also near peak phase. We see that for this proxy of the cycle, the
time of peak response during the declining phase is close to a month longer 
than during the rising phase, similar to what was found from toroidal fields 
at $15^{\circ}$ and $60^{\circ}$ seen in Figures 4 and 6. On the other hand, the 
difference in response time for flow increases and decreases has shrunk. This 
is probably because the effect of a given change in meridional flow is 
different at different latitudes, and all latitudes of the tachocline are 
included in the flux integral. By contrast, Figure 7f shows that near cycle 
maximum the difference in response time between flow increase and decrease is 
even larger. From Figure 7c, it seems clear that near maximum, speed-up or 
slow-down in phase advancement becomes convolved with changes in peak amplitude.
The higher peak from flow increase comes later than the lower peak with flow 
decrease, but this apparent phase difference is reversed as the declining phase
progresses.

\subsection{Model's response to variation in the meridional flow-speed
durations shorter and longer than 9 months}

For results presented so far, we have chosen a meridional circulation
perturbation that lasts nine months. What happens when the length of this 
perturbation is changed? Figure 8 gives the answer. Here we display the
lag correlations for meridional circulation perturbations of the same
profile but durations of 3,4,6,9,12 and 18 months. As the duration 
of perturbation is lengthened from 3 to 18 months, several features
are revealed. First, the time of peak response increases, approximately 
in proportion to the duration of the perturbation. To first order, 
the time of peak response remains close to the time width of the meridional 
flow perturbation at the point of full width at half maximum. But by 24 months, 
the response time is somewhat shorter than that value (10 months vs 12 months). 
We judge this to be a real effect, because the response time should remain 
finite in the face of a permanent change of the meridional flow speed. There 
should be an asymptotic response time in the system as the duration of the 
meridional flow perturbation is increased. We have not attempted to find this 
asymptotic value, but it is clearly longer than ten months in the 
flux-transport dynamo we have used.

Second, the difference between the time of peak response for flow increase and 
flow decrease also increases; the systematic increase in the separation 
between the two vertical pale green marks from top right frame towards 
the bottom right frame of Figure 8 is clearly evident. As seen earlier 
in Figures 4, 5, 6 and 7, the time of peak response is shorter in the case of 
flow-increase than in the case of flow-decrease. During the forward progress 
of the cycle the increase in flow helps the cycle progress faster 
and hence, the increase in flow helps transmit the change faster to 
neighboring points of the domain than that in the case of flow decrease. 
The more enhanced is this effect, the longer is the duration of flow 
perturbation. 

We compare the times of peak flow-perturbation and peak response for
a wide range of perturbation durations in Figure 9. The left hand scale
gives both peak times in months, while the right hand scale is a 
measure of the difference between these times, normalized by the duration
of the flow perturbation (twice the time to the peak of the flow-
perturbation). We see that for long peak times, their normalized difference
approaches zero, i.e., the model responds at about the same rate as the applied
perturbation. But for short durations, the normalized difference reaches
0.5, meaning the time of peak response is twice as long as the time to
peak perturbation. It appears this ratio will get even larger for 
perturbations of even shorter duration; this is pushing the limit of 
applicability of a mean-field formulation. Nevertheless, by extrapolation 
of the plots of peaks to zero duration, it appears that the shortest possible 
peak response time is about 1.5 months. But as we have shown, for a more
typical meridional flow perturbation of 3-18 months, the response is spread 
over a much longer time period.

\clearpage
\begin{figure}[hbt]
\epsscale{0.75}
\plotone{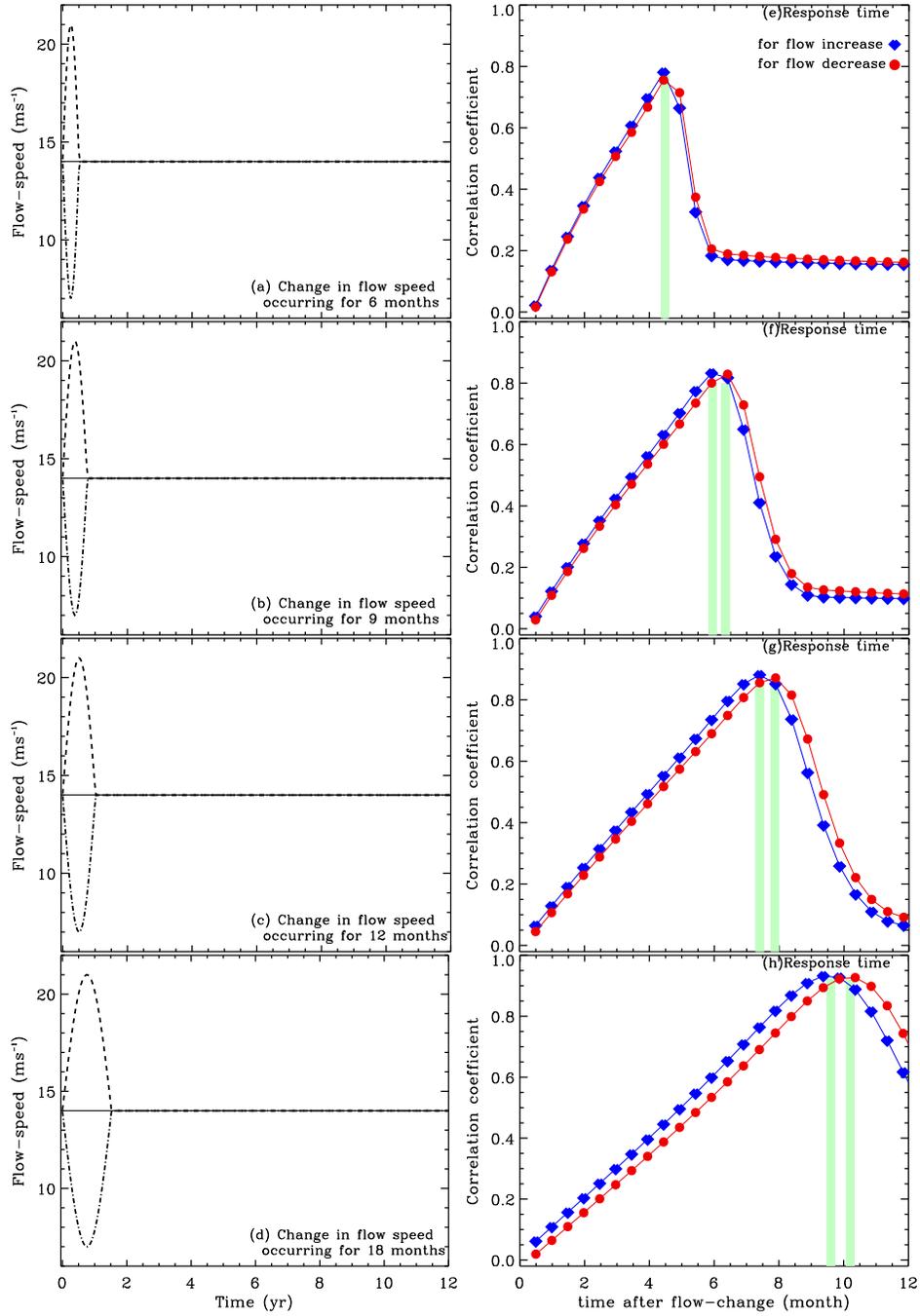}
\caption{
Four frames from top to the bottom in left column present flow-changes 
with different durations, respectively for 6, 9, 12 and 18 months. 
Corresponding lag-correlation coefficients between total tachocline
toroidal flux and flow-variation have been plotted in frames (e-h)
in right column. 
}
\label{flux_response_long}
\end{figure}
\clearpage

\begin{figure}[hbt]
\epsscale{1.0}
\plotone{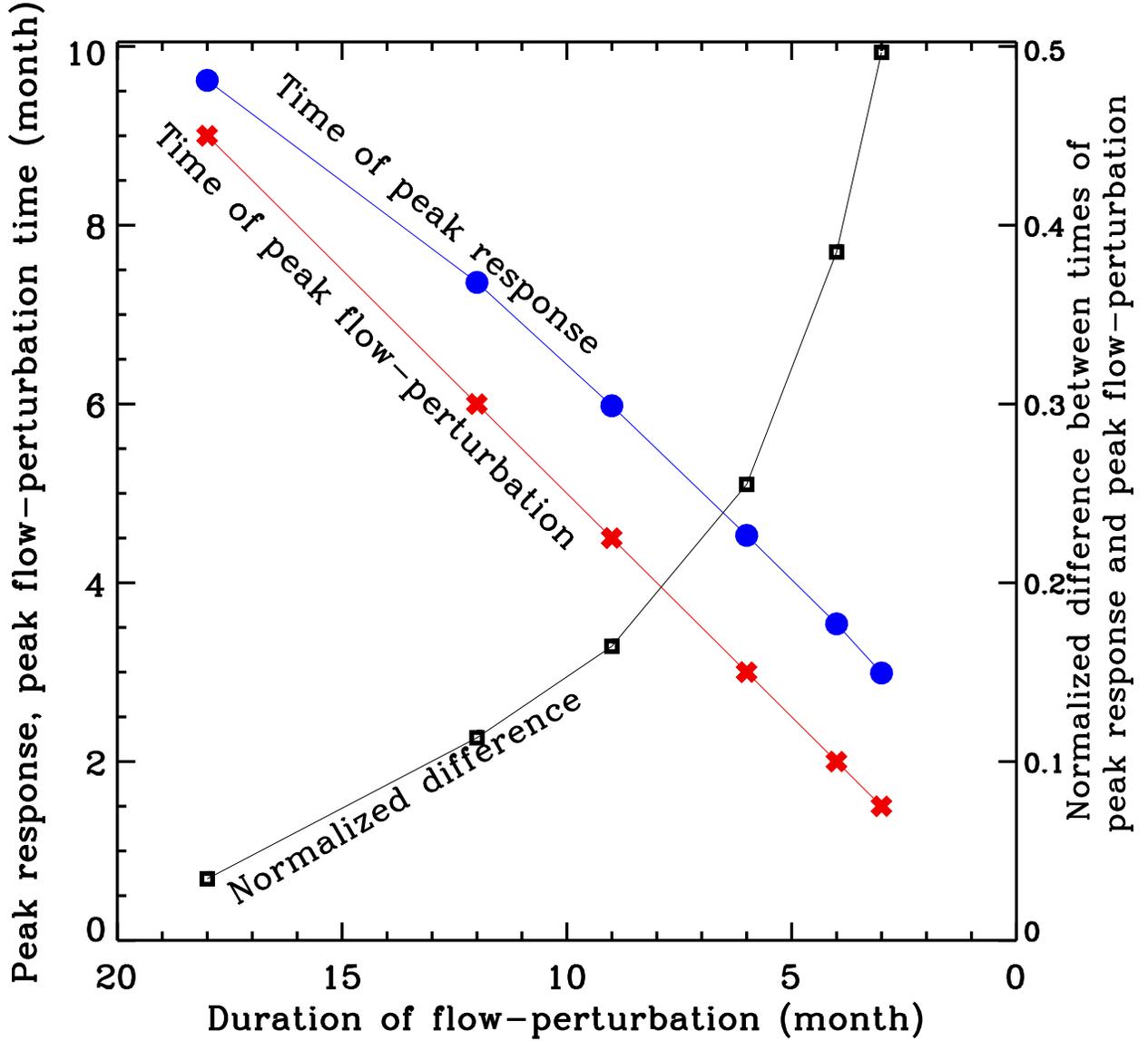}
\caption{Comparison of times of peak flow perturbation and peak response
}
\label{time comparison}
\end{figure}

The shortest possible peak response time should be greater than zero,
because it takes a certain time for the effect of the changed meridional
circulation amplitude to be transmitted from grid point to the next. Since
our calculations are for an advection-dominated dynamo, advection times
will be shorter than diffusion times, but all ill contribute. We used
101 grid points between inner and outer boundaries, and the same between
equator and pole. Therefore, the transmission time is four times longer
between adjacent points in latitude than between adjacent radial points,
for the same velocity. But mass continuity, the latitudinal flow is 
$\sim 4$ or 5 times larger than the radial flow; therefore the advection
times between adjacent grid points are roughly the same in radial and 
latitudinal directions. For example, for an average latitudinal flow
speed of $4\,{\rm m}{\rm s}^{-1}$ and radial speed of $1\,{\rm m}{\rm s}^{-1}$,
the transmission time is about 23 days. We should expect the peak
response to be reached in a somewhat longer time than 23 days; we found
it roughly to be 1.5 months in the present calculations.

\subsection{Model's response to variation in the meridional flow-speed
in the case of a higher diffusivity}

Another physical parameter that the time of peak response could be sensitive 
to is the magnetic diffusivity. Figure 10 displays the time variation in 
toroidal flux integral and the lag correlation for a meridional flow 
perturbation for nine months, applied during the rising phase of a dynamo 
solution with the magnetic diffusivity of the bulk of the dynamo domain 
doubled to $10^{11}\,{\rm cm}^2\,{\rm s}^{-1}$. We see that the time of peak response  
remains about six months, essentially the same as found for the lower 
magnetic diffusivity case shown in Figure 7d. We infer from this result 
that as long as the dynamo is operating in the advection-dominated 
regime, the response time does not depend significantly on the diffusivity. 
For the dynamo model we have used, solutions with magnetic diffusivity 
smaller than about $2\times 10^{11}\,{\rm cm}^2\,{\rm s}^{-1}$ are 
advection dominated. For substantially higher diffusivities, when 
diffusion dominates over advection, we would expect the time of peak response 
time to be influenced by the magnetic diffusivity value. We have not 
explored that regime here.

\clearpage
\begin{figure}[hbt]
\epsscale{0.65}
\plotone{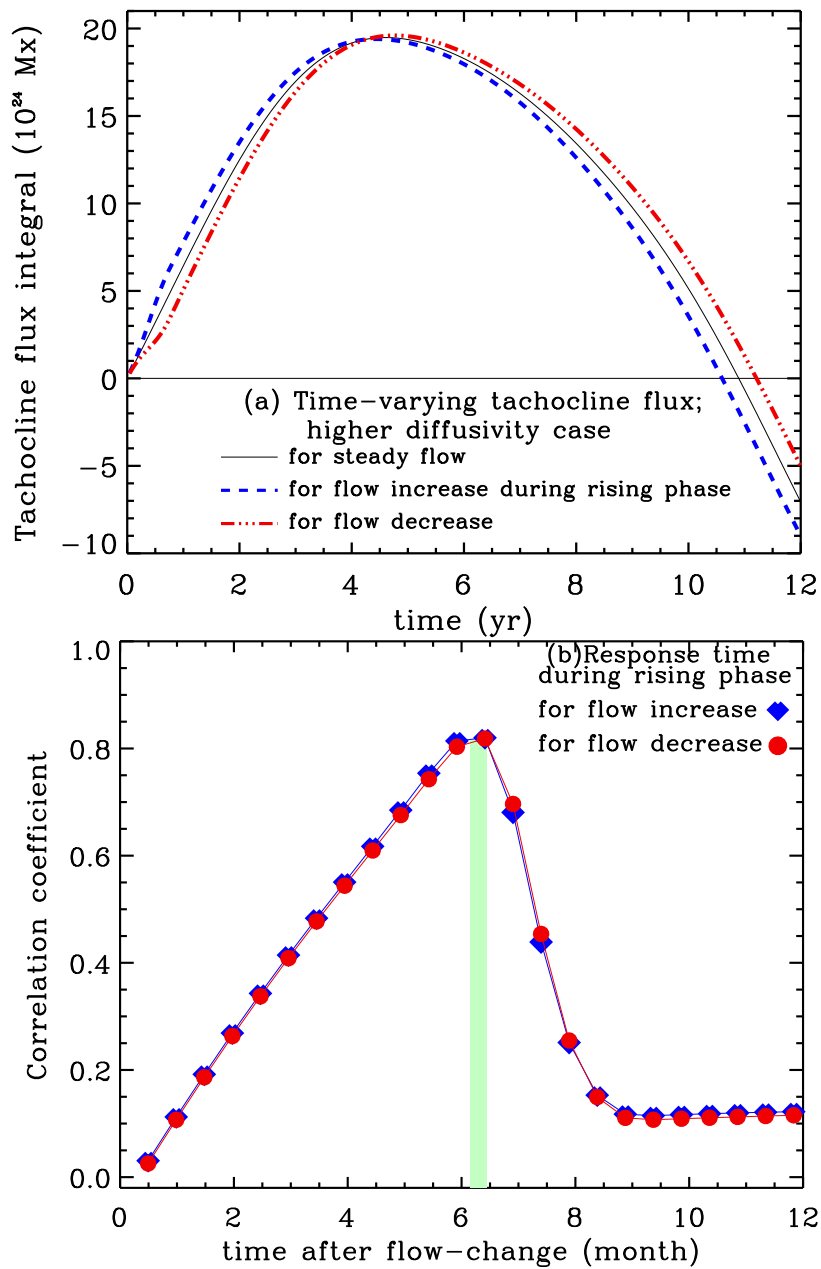}
\caption{Toroidal flux integral and lag correlation profiles for a dynamo
solution with double the magnetic diffusivity of solutions displayed in 
Figure 6. Meridional flow perturbation lasts 9 months and is applied during
the rising phase of the cycle}
\label{flux_response_higheta}
\end{figure}
\clearpage

\subsection{Model's response to a complex variation in the meridional 
flow-speed}

So far, we have done all of our numerical experiments performed
with one particular profile of flow perturbation, namely a $\pm \sin(\Delta
\tau)$ type profile for flow change with respect to the steady, mean flow.
In this section, we perform two experiments to investigate how the
model's response time depends on the choice of flow perturbation profile
and amplitude. In Figures 11(b) and (c), we present the time variation of
toroidal flux integral and its lag correlation for a meridional flow
perturbation as shown in Figure 11a. The duration of the flow perturbation 
is 24 months in this case; during the first 6 months the flow increases 
by 50\%, and then continuously decreases during the next 12 months, so 
that the speed gets reduced by 50\% with respect to the steady, mean 
flow of $14\,{\rm m}\, {\rm s}^{-1}$, and finally it increases again 
during the last 6 months of 24 months' perturbation to reach the level 
of steady, mean flow speed. 

The toroidal flux integral (TFI) in Figure 10b in the case of time-varying 
flow shows a fast phase advancement with respect to that
for steady flow (thin black curve) and then a slow-down followed by a 
slight speed-up. The two TFI patterns (dashed blue and thin black curves)
ultimately match in the late declining phase of the cycle. This is not 
surprising, because the time-averaged flow-speeds are the same in both cases 
(see Figure 11a). However, what is surprising is that the time of peak
response in this case with more complex flow perturbation than that used in
Figures 4-9 is again about 6 months (see Figure 11c), which is within 
a similar range of response times we found in \S3.1, \S3.2 and \S3.3. 
 
Although a flow perturbation in step-function form may not be a realistic
flow variation in the case of solar meridional circulation, we show two
cases to study the model's response to flow variation in step-function
profiles. Selecting two different durations of such flow variations, namely
the perturbations lasting for 6 months and 1 year respectively (see blue-dashed
and bluish green dash-dotted lines in Figure 12(a)) during the rising phase of 
a solar cycle, we compute the TFI and the model's response to the above two 
flow perturbations and plot in Figures 12(b) and (c) respectively. Again 
blue-dash and bluish-green dash-dotted lines represent respectively the
TFIs for flow perturbations with 6 months and 1 year duration. 
The response times of the model for these two cases of step-function
flow perturbations are 5.4 months and 8.6 months respectively. This
experiment gives another example of similar lag correlation and model's
response time as shown in Figures 4-9, indicating the robustness of a
dynamo model's response to meridional flow variation. The peak correlation
is lower for a stepfunction of 12 months duration than for a sinusoidal
perturbation of the same duration, because the flow perturbation peak
itself is flat for a year, rather than strongly peaked.

\begin{figure}[hbt]
\epsscale{0.5}
\plotone{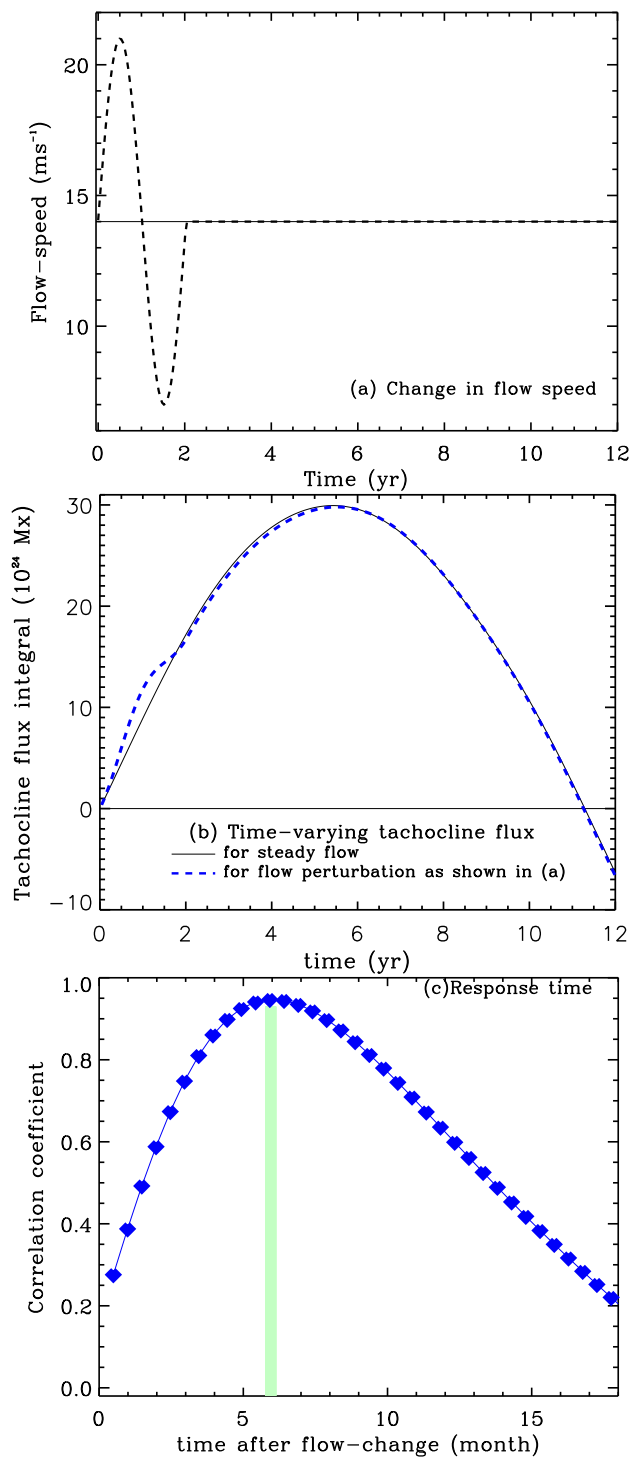}
\caption{Frames (a,b,c) respectively present flow variation, toroidal 
flux integral and lag correlation profiles for a dynamo solution with a 
different and more complex flow perturbation than that in Figure 4(a).
Meridional flow perturbation lasts for two years and is applied during 
the rising phase of the cycle}
\label{flux_response_complex}
\end{figure}

\clearpage
\begin{figure}[hbt]
\epsscale{0.5}
\plotone{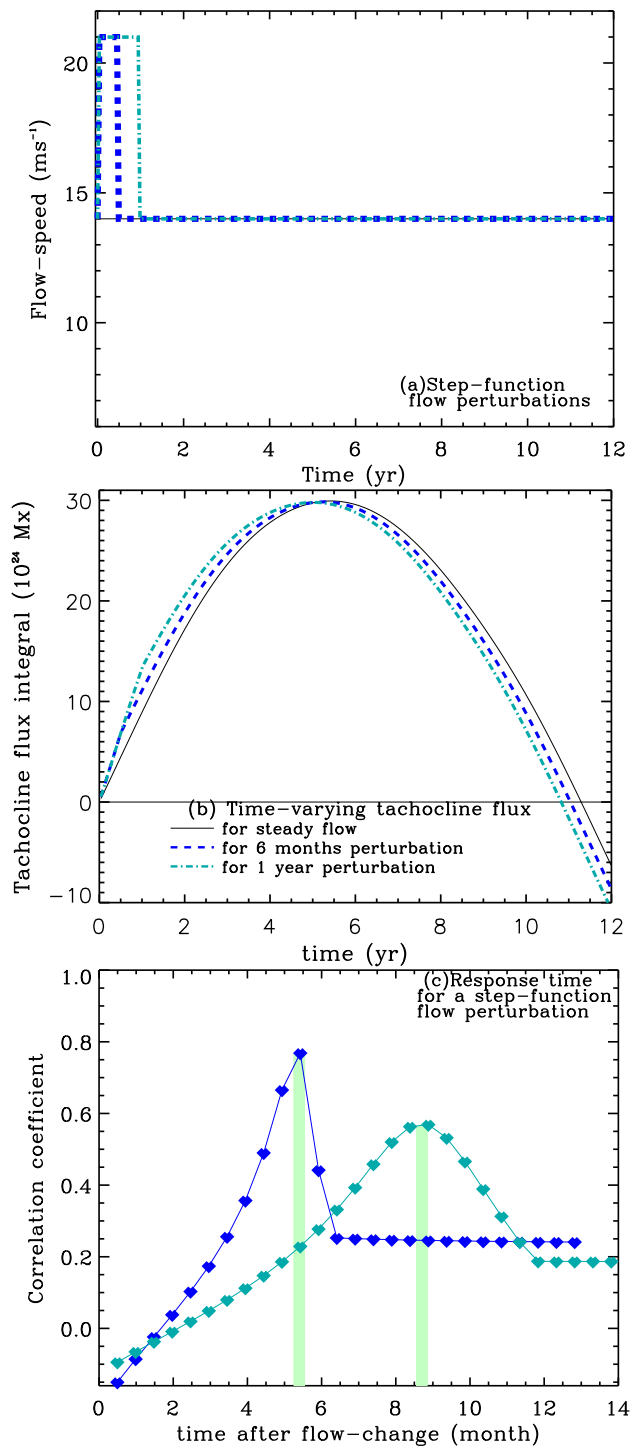}
\caption{Top frame presents step-function flow variation with durations
of 6 months (blue-dashed line ) and 1 year (bluish-green dash-dotted line).
Corresponding toroidal flux integrals and response times of the model
are shown in frames (b) and (c) respectively.}
\label{flux_response_step}
\end{figure}
\clearpage

In order to investigate whether the amplitude of the flow perturbation 
can influence the model's response to flow variation we perform dynamo
simulations with different amplitudes of flow perturbation, namely
25\%, 50\% and 100\% changes in flow speed with respect to the steady flow
of $14\,{\rm m}\, {\rm s}^{-1}$. In the previous sections, we focused 
only on the model's response to 50\% change in flow speed. 
In Figures 13(a), (b) and (c), we respectively show the flow perturbation, 
the TFI and the response time. Flow perturbations with 25\%, 50\% and 100\%
amplitudes, compared to steady flow (thin black line in Figure 11a), 
are shown in green long-dashed, blue dash-dotted and black short-dashed
lines. If we had included a case for which the flow amplitude was increased
by 75\%, this curve would have fallen in between the 50\% and 100\%
curves. Corresponding TFIs in Figure 11b show that the changes in phase
advancement are approximately proportional to the amplitudes of flow
perturbations. Consequently lag correlation patterns, plotted in Figure 11c 
in small-sized green, medium-sized blue and large black diamonds respectively
for 25\%, 50\% and 100\% amplitudes of flow perturbations, almost overlap 
with one another. The time of peak response is 7.4 months, which indicates
again that the model responds to the flow perturbation according to
the inherent memory of the model \citep{ynm2008}. The model's time of peak
response is not influenced by the amplitude of the flow perturbations because
the the rate of cycle phase changes roughly in proportion to the rate
of change of flow amplitude in this class of models. 

\clearpage

\begin{figure}[hbt]
\epsscale{0.45}
\plotone{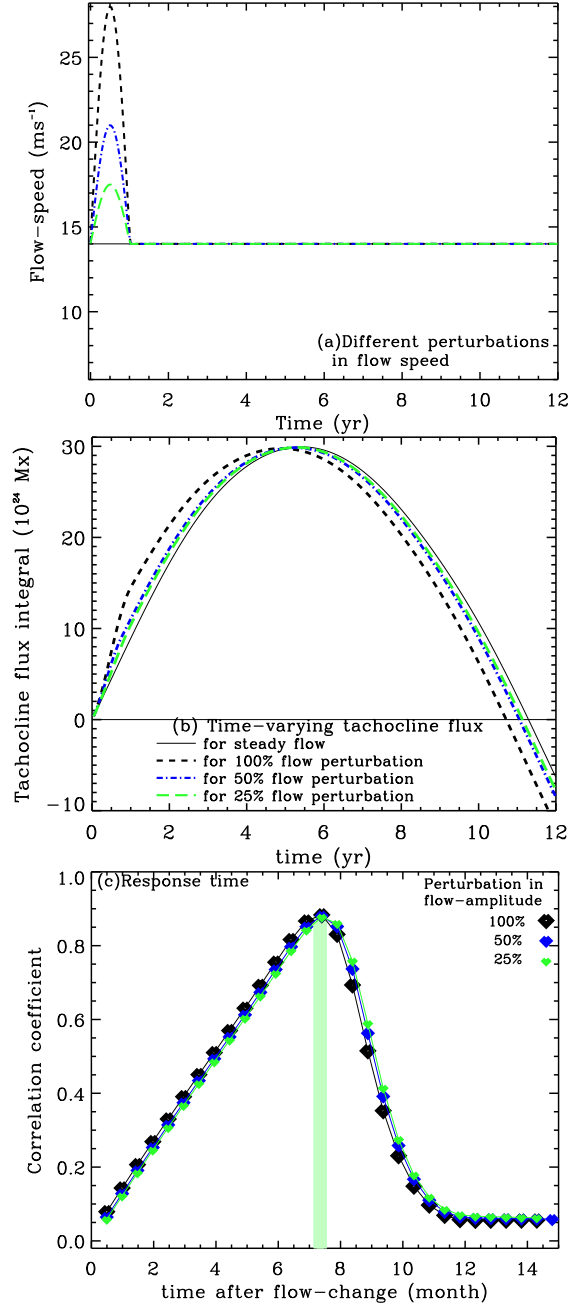}
\caption{Green dashed, blue dash-dotted and black long-dash lines in 
frame (a) present three different amplitudes of flow perturbations, 
25\%, 50\% and 100\% with respect to steady flow ($14\,{\rm m}\,{\rm s}^{-1}$). 
In frame (b) the corresponding curves present toroidal flux integrals 
and in frame (c) small-sized green, medium-sized blue and large black diamonds 
present lag correlation plots for the above three flow perturbations
respectively. Flow perturbation lasts for 1 year and occurs during the
rising phase of the cycle.}
\label{flux_response_himidlo}
\end{figure}
\clearpage

\section{Discussion and Conclusions}

Solar cycle prediction research has progressed significantly during
recent years so that physics-based models are being integrated 
forward in time, in addition to using empirical relations. But 
so far those efforts have been limited to predictions of peak-amplitude 
of a cycle and its average duration. Attempts to predict simultaneously 
the shape, timing and amplitude of a cycle have not been made yet. With 
knowledge gained from advances in predictions employing oceanic and 
atmospheric climate models, it has proven necessary to move beyond 
simple data-nudging to more sophisticated data-assimilation 
techniques. Efforts using an Ensemble-Kalman Filter (EnKF) approach 
\citep{kk09} as well as variational approach \citep{jbt11} have 
been made using one-dimensional dynamo models. 

Prior results from flux-transport dynamo simulations reveal that the
meridional circulation primarily governs the progress of a cycle's
phase in this particular class of dynamo models. The time variations 
in meridional circulation speed and profile can be used in a sequential
data-assimilation approach that involves an EnKF method in the framework 
of the Data Assimilation Research Testbed (DART) \citep{a09} in order to 
predict the details within a cycle, namely the rise and fall patterns, 
the onset, peak and end timings and the peak-amplitude. 

A sequential data assimilation approach can be most efficiently used 
if we know the model's response time to a change in its ingredients. 
Then we will know how to update the input data of a specific ingredient
into the model. With this motivation we studied a flux-transport
dynamo model's response time to meridional flow-speed variations, and
found that the model's time of peak response to a change in flow-speed 
that lasts typically from one-half to one and a half years, is about six 
months on average. This response time is independent of proxies used to 
measure a theoretical solar cycle, such as tachocline toroidal field 
at $15^{\circ}$ latitude, at $60^{\circ}$ latitude or the 
integrated total toroidal flux in the tachocline. All response times
are much shorter than the 'circulation time' or the time it takes for
a fluid element to make a complete circuit on a closed streamline that
reaches both low and high latitudes as well as passing close to the
inner and outer radial boundaries of the dynamo domain.

Incorporating changes in flow-speed lasting for different time-spans,
such as 3, 6, 9, 12 and 18 months, we found that the time of peak response 
of the model increases with the duration of flow variation; it
is approximately the length of full width at half maximum of
the flow-perturbation profile in time. Response of the model is found
to be always slightly faster when the flow change is positive with
respect to the mean, steady flow, primarily because the rate of
progress in a cycle's phase is approximately proportional to the
flow-speed in this class of dynamo models. We also found that the time of 
peak response is independent of magnetic diffusivity so long as the dynamo 
operates in the advection-dominated regime. 

Although in most of our numerical experiments we chose a smooth 
sinusoidal type variation in flow-speed, we have also performed 
numerical experiments incorporating different shape and amplitudes
of flow perturbations. We found that the model's time of peak response
to change in flow is roughly independent of the shape and 
amplitude of the flow perturbations.

There exists observations of systematic decrease of flow speed 
during the entire rising phase of the cycle 23 \citep{ba2003}.
Consideration of the flow speed change during such a long span of
time for studying the response time of a dynamo models is beyond
the scope of this paper. In a sequential assimilation scheme 
the unknown model ingredients require updating more frequently
than on a solar cycle time scale in a dynamo model that is attempting
to simulate shape of a solar cycle. So the experiments we have 
performed here have the perturbation lasting for no more than
two years. However, it will be interesting to investigate whether 
a dynamo model would respond within the same cycle or in the next 
cycle to a systematic flow variation that occurs during an entire 
rising or declining phase.

We have focused only on the speed variation and taken no variation 
in latitudinal or radial flow profile, because we do not have 
information from observations about the complete flow-profiles in 
the convection zone. We can implement the knowledge gained here 
about the model's response time to change in flow-speed to develop 
EnKF schemes for assimilating time-varying flow data sequentially 
and simulate a cycle's rise and fall patterns along with its 
amplitude and timing.

Flux-transport dynamo models are particularly sensitive to a 
meridional flow changes; so its influence may be overestimated 
in this class of models. Data assimilation techniques are useful 
to better determine the relevant ingredients in the mean-field 
dynamo models and quantify their influence on observables like 
the magnetic cycle period.

\acknowledgements

We thank Peter Gilman for reviewing the entire manuscript and for
many helpful discussions. We also thank an anonymous reviewer
for many helpful comments and constructive criticism which have
helped significantly improve this paper. We extend our thanks to
Sacha Brun for suggesting experiments with different amplitudes 
of flow perturbations. We acknowledge the support from NORDITA 
school on 'predictability and data assimilation' -- various 
discussions in that school have helped us formulate and solve this 
problem presented in this paper. This work is partially supported 
by NASA grant NNX08AQ34G. The National Center for Atmospheric Research
is sponsored by the National Science Foundation. 

\clearpage
%%%%%%%%%%%%%%%%%%%%%%%%%%%%%%%%%%%%%%%%%%%%%%%%%%%%%

\end{document}